# Data Science vs Putin: How much does each of us pay for Putin's war?

Fabian Braesemann, Max Schuler

4. March 2022

## Abstract

Putin's Ukraine war has caused gas prices to skyrocket. Because of Europe's dependence on Russian gas supplies, we all pay significantly more for heating, involuntarily helping to fund Russia's war against Ukraine. Based on an analysis of real-time gas price data, we present a calculation that estimates every household's financial contribution for heating paid to Russian gas suppliers daily at current prices - six euros per household and day. We show ways everyone can save energy and help reduce the dependency on Russian gas supply.

## Introduction

People in Europe are shocked to see the images of severe destruction and human suffering in Putin's Ukraine war. Western sanctions are already having an impact on the Russian economy. However, one of Russia's most important sources of foreign income is not affected by the sanctions: exports of gas, oil and raw materials. This is why the public debate about a possible embargo of Russian gas imports has increased in recent days [1].

Such a step would most likely lead to further increases in energy prices, because of Europe's high dependency on Russian gas imports. These high and short-term costs for households and the economy will exist until alternatives are made accessible.

Because of the current high gas prices, a lot of money of goes to Russian gas suppliers if we turn on the heating and use hot wat. But how much exactly does every one of us pay for Putin's war? And what can we do to reduce this amount?

Based on online data about recent gas price developments, we present an assumption-based calculation that estimates the daily financial contribution of individual households in Germany to Russian gas suppliers and we show savings potential.
The analysis presented here uses data science methods. Data science, in particular the structured collection and analysis of alternative data sources (especially online data such as the data on gas price developments used here), offers the possibility of empirically analysing a large number of social and economic phenomena with high temporal resolution and to describe them in near real-time. Data science methods can thus provide evidence-based insights into socially relevant topics in times of rapidly changing conditions.

## The spot market price for gas delivery has doubled since mid-February

Gas prices skyrocketed since the start of the Ukraine war. Figure 1 shows the development of THE (Trading Hub Europe) spot market prices for gas delivery in Germany from January 18 to March 3, 2022.

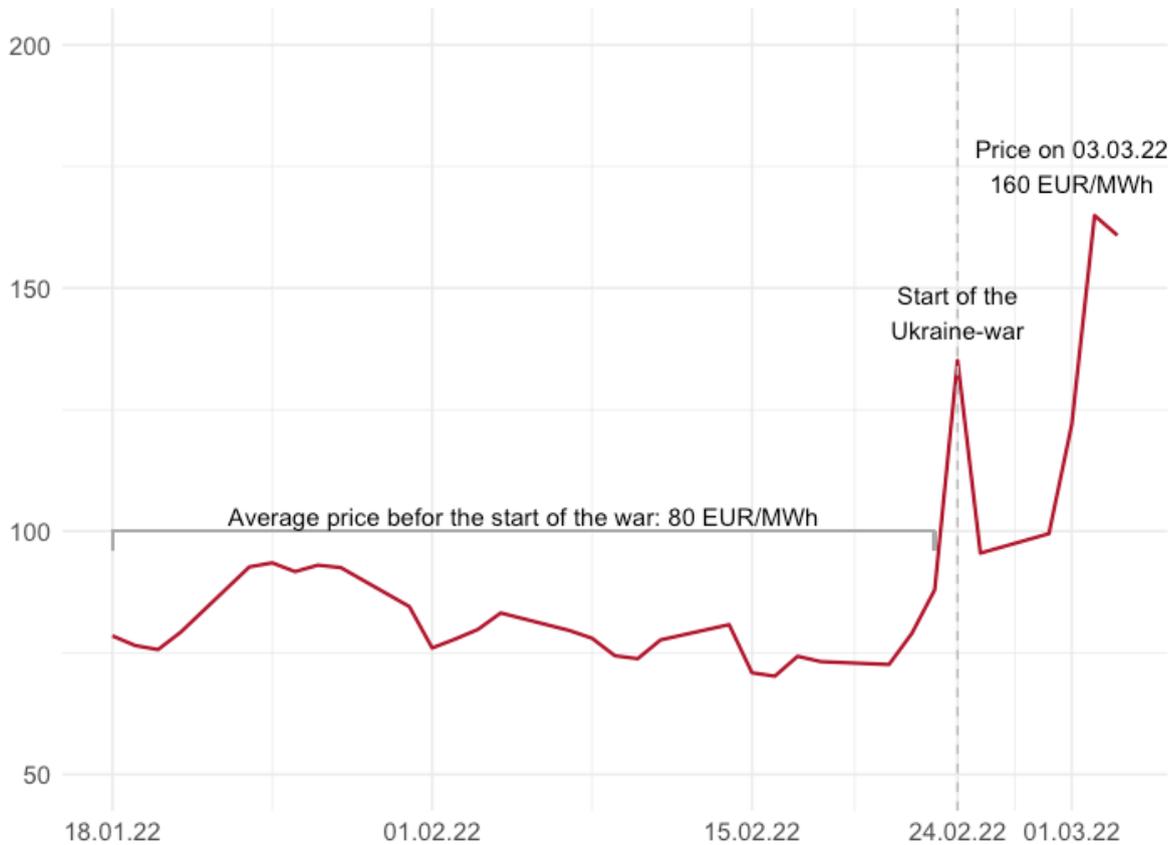

*Figure 1 - Development of THE spot market prices for Germany from January 18 to March 3, 2022: Prices have doubled since the beginning of the war.*

While the THE spot market price per megawatt-hour (MWh) averaged on around €80 in the weeks before the war began, the current price as of March 3 is €160 per MWh [2]. That means, while each and every one of us made a significant contribution to Putin's finances by heating already before the start of the war due to the high gas prices, now every day of heating flushes twice as much into the Russian war chest. The THE forward prices for the coming months and quarters show a similarly exposed price level.

How much money do German households pay for Russian gas in total?

With the current very high prices for natural gas, there is a significant amount of money that all German households with gas heating have to pay for Russian gas every day.

Table 1 gives an overview of the total daily payments of all German households for Russian gas. Almost half (48%) of the 42.5 million German homes are heated with gas. An average-sized apartment (92 m²) needs around 72 kWh of energy from gas for an average heating day. This results in total daily consumption of around 1.47 terawatt-hours (TWh) for an average heating day.

| Daily payments for Russian gas supply used for heating (all German households) | |
|---|---|
| Number of apartments in Germany [3] | 42.5 Mio. |
| Average size of apartments | 92 Quadratmeter |
| Share of apartments with gas heating [4] | 48 % |
| Daily gas consumption of a 92qm apartment | 72 kWh |
| Daily gas consumption of all German households | 1.47 TWh |
| Daily payments to Russian gas suppliers (50 % of all gas consumption) | €117 Mio. |

*Table 1 - Daily payments of all German households for Russian gas deliveries for heating with natural gas at current gas prices:* Assuming average values, the daily payments to Russia amount to 117 million euros.

As around half of the natural gas consumed in Germany comes from Russia, current market prices result in daily payments of €117 million for this consumption. This means that German households pay more than one hundred million euros to Russian gas suppliers for an average day of heating, thereby helping to finance Putin's war in Ukraine.

### What every one of us can do

What do these numbers mean for every one of us? Table 2 breaks the total numbers down to the level of individual apartments of different sizes. An average household in Germany needs 72 kWh of energy per heating day. At the current market prices for gas, this results in €5.72 that every household with gas heating contributes to Putin's war chest every day.

In other words, every household pays almost six euros per heating day to Russian gas suppliers. Anyone with a gas-fired heater helps fund Russia's war by consuming gas delivered from Russia.

Contribution of individual households and individual savings potential
(Assumptions: 180 heating days per year, 50 % gas supplied from Russia,
Savings of 12 % through room temperature reduction by 2°C [5])

| Household size | Daily gas consumption [6] | Daily payments to Russia | Daily savings through 2°C temperature reduction |
|---|---|---|---|
| 40 m² | 31 kWh | €2.49 | €0.30 |
| 60 m² | 47 kWh | €3.73 | €0.45 |
| 80 m² | 62 kWh | €4.98 | €0.60 |
| 100 m² | 78 kWh | €6.22 | €0.75 |
| 120 m² | 93 kWh | €7.47 | €0.90 |
| **Values of an average household in Germany** | | | |
| 92 m² | 72 kWh | €5.72 | €0.69 |

*Table 2 - Average daily payments by individual households during the heating season for Russian gas deliveries for heating with natural gas at current gas prices and individual savings potential:* At current gas prices, an average household in Germany pays €5.72 per day to Russian gas suppliers during the heating season. A temperature drop of 2°C could reduce this amount by €0.69 per day.

What you can do to reduce this involuntary subsidy?
Until large-scale alternatives to Russian gas imports are found, reduction of consumption by households can be an effective means to reduce spending on Russian gas.

As Table 2 shows, reduced gas consumption by lowering the average room temperature by 2°C leads to an average reduction in Russian income of €0.69 per average heating day and household. If all 20.5 million German households with gas heating lower the room temperature accordingly, this would lead to a reduction in income of more than 14 million euros for Russian gas deliveries per average heating day. This amount could even be higher if all households in Europe lower the room temperature accordingly: our calculation shows that up to 70 million euros could be saved per day in that way.

Over the remaining duration of the heating period until the end of March, this could result in savings of more than 380 million euros by German households (or 1.8 billion euros if all European households were to lower the room temperature by 2°C). This is a significant amount of money that is not being made available to the Russian state and its war. In addition, long-term climate-friendly investments such as energy-saving refurbishments can help to lower gas consumption permanently.

While we cannot directly alleviate the suffering of the people in Ukraine from abroad, each of us can do something. In addition to humanitarian aid and donations, we can contribute through individual savings to prevent Putin from being rewarded for his warmongering with higher payments for gas supplies.

Together we can help to reduce the dependency on Russian gas supplies by using heating energy consciously and reducing consumption. Two degrees less room temperature in the living room can easily be dealt with in wearing a sweater at home or sitting on the sofa in a blanket, but already make a substantial contribution.

If you want to do more and are hardened, you can go one step further to reduce your own gas consumption: cold showers. 550 kWh of energy per year are used per person to heat the daily 40 litres of water with gas that are needed on average for showering [7]. This results in a consumption of 1.5 kWh of energy or €0.12 to Russian gas suppliers per shower, which can be avoided by taking a cold shower.

So, grit your teeth and take a cold shower. The positive side effect? Cold showers are said to make you fit, happy and beautiful [8]. And it's also good for the climate.

### About us

Dr Fabian Braesemann is a Research Fellow & Data Scientist at the Saïd Business School at the University of Oxford and a Research Associate at the Oxford Internet Institute and the Humboldt Institute for Internet and Society Berlin. He is also the founder of the Datenwissenschaftliche Gesellschaft (DWG) Berlin, a company that uses data science methods to solve social and economic problems.

Max Schuler is Global Category Manager in the Power & Natural Gas division at HeidelbergCement AG. He has many years of experience in the field of energy procurement and energy solutions for multinational corporations.